\documentclass[12pt]{article}
\usepackage{graphicx}
\usepackage{color}
\usepackage{authblk}
\def\hybrid{\topmargin 0pt      \oddsidemargin 0pt
        \headheight 0pt \headsep 0pt
       \voffset-1cm
        \textwidth 6.25in       % A4 paper
       \textheight 9.5in       % A4 paper
        \marginparwidth 0.0in
        \parskip 5pt plus 1pt   \jot = 1.5ex}
\catcode`\@=11
\def\marginnote#1{}

\newcount\hour
\newcount\minute
\newtoks\amorpm
\hour=\time\divide\hour by60
\minute=\time{\multiply\hour by60 \global\advance\minute by-\hour}
\edef\standardtime{{\ifnum\hour<12 \global\amorpm={am}%
        \else\global\amorpm={pm}\advance\hour by-12 \fi
        \ifnum\hour=0 \hour=12 \fi
        \number\hour:\ifnum\minute<10 0\fi\number\minute\the\amorpm}}
\edef\militarytime{\number\hour:\ifnum\minute<10 0\fi\number\minute}

\def\draftlabel#1{{\@bsphack\if@filesw {\let\thepage\relax
   \xdef\@gtempa{\write\@auxout{\string
      \newlabel{#1}{{\@currentlabel}{\thepage}}}}}\@gtempa
   \if@nobreak \ifvmode\nobreak\fi\fi\fi\@esphack}
        \gdef\@eqnlabel{#1}}
\def\@eqnlabel{}
\def\@vacuum{}
\def\draftmarginnote#1{\marginpar{\raggedright\scriptsize\tt#1}}

\def\draftlabel#1{{\@bsphack\if@filesw {\let\thepage\relax
   \xdef\@gtempa{\write\@auxout{\string
      \newlabel{#1}{{\@currentlabel}{\thepage}}}}}\@gtempa
   \if@nobreak \ifvmode\nobreak\fi\fi\fi\@esphack}
        \gdef\@eqnlabel{#1}}
\def\@eqnlabel{}
\def\@vacuum{}
\def\draftmarginnote#1{\marginpar{\raggedright\scriptsize\tt#1}}

\def\draft{\oddsidemargin -.5truein
        \def\@oddfoot{\sl preliminary draft \hfil
        \rm\thepage\hfil\sl\today\quad\militarytime}
        \let\@evenfoot\@oddfoot \overfullrule 3pt
        \let\label=\draftlabel
        \let\marginnote=\draftmarginnote
   \def\@eqnnum{(\theequation)\rlap{\kern\marginparsep\tt\@eqnlabel}%
\global\let\@eqnlabel\@vacuum}  }

\def\numberbysection{\@addtoreset{equation}{section}
        \def\theequation{\thesection.\arabic{equation}}}

\def\underline#1{\relax\ifmmode\@@underline#1\else
        $\@@underline{\hbox{#1}}$\relax\fi}

\def\titlepage{\@restonecolfalse\if@twocolumn\@restonecoltrue\onecolumn
     \else \newpage \fi \thispagestyle{empty}\c@page\z@
        \def\thefootnote{\fnsymbol{footnote}} }

\def\endtitlepage{\if@restonecol\twocolumn \else  \fi
        \def\thefootnote{\arabic{footnote}}
        \setcounter{footnote}{0}}  %\c@footnote\z@ }
\relax

\numberbysection
\hybrid

\newfont{\Bbb}{msbm10 scaled 1\@ptsize00}
\newfont{\Bbbb}{msbm7 scaled 1\@ptsize00}

\newcommand{\DDD}{\raise-1pt\hbox{$\mbox{\Bbbb D}$}}

\newcommand{\UUU}{\raise-1pt\hbox{$\mbox{\Bbbb U}$}}

\newcommand{\z}{\raise-1pt\hbox{$\mbox{\Bbbb Z}$}}

\def\res{\mathop{\hbox{res}}\limits}

\def\beq{\begin{equation}}
\def\eeq{\end{equation}}
\def\p{\partial}

\begin{document}

\begin{titlepage}

\title{Elliptic solutions to 
Toda lattice hierarchy and elliptic Ruijsenaars-Schneider model}

\author[1,2]{V.~Prokofev\thanks{vadim.prokofev@phystech.edu }}
\author[2,3,4]{
 A.~Zabrodin\thanks{ zabrodin@itep.ru}}
 \affil[1]{Moscow Institute of Physics and Technology, Dolgoprudny, Institutsky per., 9,
Moscow region, 141700, Russia}
 \affil[2]{
Skolkovo Institute of Science and Technology, 143026 Moscow, Russian Federation
}
\affil[3]{ National Research University Higher School of Economics,
20 Myasnitskaya Ulitsa, Moscow 101000, Russian Federation}
\affil[4]{
ITEP NRC KI, 25 B.Cheremushkinskaya, Moscow 117218, Russian Federation
}

%\vspace{-5cm} \centerline{ \hfill ITEP-TH-17/19}\vspace{7cm}

\date{February 2021}
\maketitle

\vspace{-11cm} \centerline{ \hfill ITEP-TH-06/21}\vspace{11cm}

\begin{abstract}

We consider solutions of the 2D Toda lattice hierarchy which are elliptic 
functions of the ``zeroth'' time $t_0=x$. It is known that their poles as functions
of $t_1$ move as
particles of the elliptic Ruijsenaars-Schneider model. The goal of this paper 
is to extend this correspondence
to the level of hierarchies. We show that
the Hamiltonians which govern the dynamics of poles with respect to the $m$-th hierarchical
times $t_m$ and $\bar t_m$ of the 2D Toda lattice hierarchy 
are obtained from expansion of the spectral curve for the Lax matrix of the 
Ruijsenaars-Schneider model at the marked points. 

\end{abstract}

\end{titlepage}

\tableofcontents

\vspace{5mm}

\section{Introduction}

\subsection{Motivation and result}

The 2D Toda lattice (2DTL) hierarchy  \cite{UT84} is an infinite set
of compatible nonlinear dif\-fe\-ren\-ti\-al-\-dif\-fe\-rence 
equations involving infinitely many
time variables ${\bf t}=\{t_1, t_2, t_3, \ldots \}$ (``positive'' times), 
$\bar {\bf t}=\{\bar t_1, \bar t_2, \bar t_3, \ldots \}$ (``negative'' times)
in which the equations
are differential and the ``zeroth'' time 
$t_0=x$ in which the equations are difference. 
Among all solutions to these equations,
of special interest are solutions which have a finite number of poles 
in the variable $x$ in a fundamental domain of 
the complex plane. In particular, one can consider solutions
which are elliptic (double-periodic in the complex plane) 
functions of $x$ with poles depending on the times.

The investigation of dynamics of poles of singular solutions to nonlinear integrable
equations was initiated in the seminal paper \cite{AMM77}, where elliptic and rational 
solutions to the Korteweg-de Vries and Boussinesq equations were studied. It was shown
that the poles move as particles of the integrable Calogero-Moser many-body system
\cite{Calogero71,Calogero75,Moser75,OP81}
with some restrictions in the phase space. As it was proved in \cite{Krichever78,CC77},
this connection becomes most natural for the more general 
Kadomtsev-Petviashvili (KP) equation, in which case there are no restrictions in the phase space 
for the Calogero-Moser dynamics of poles. 
The method suggested by Krichever \cite{Krichever80} for elliptic solutions 
of the KP equation consists in substituting the solution not in the KP equation itself but
in the auxiliary linear problem for it (this implies a suitable pole ansatz for the wave
fuction). This method allows one to obtain the equations of motion together with their
Lax representation.

Later, Shiota has shown \cite{Shiota94} that 
the correspondence between rational solutions to the KP equation and the Calogero-Moser
system with rational potential can be extended to the level of hierarchies. 
The evolution of poles with respect to the higher times $t_k$ 
of the KP hierarchy was shown to be governed by 
the higher Hamiltonians $H_k=\mbox{tr}\, L^k$
of the integrable Calogero-Moser system, where $L$ is the Lax matrix. 
More recently this correspondence was generalized to trigonometric
and elliptic solutions of the KP hierarchy (see \cite{Haine07,Z19a} for trigonometric
solutions and \cite{PZ21} for elliptic solutions). 

Dynamics of poles of elliptic solutions to the 2DTL and 
modified KP (mKP) equations was 
studied in \cite{KZ95}, see also \cite{Z19}. 
It was proved that the poles move as particles of the 
integrable Ruijsenaars-Schneider many-body system \cite{RS86,Ruij87} 
which is a relativistic generalization
of the Calogero-Moser system. The extension to the level of hierarchies
for rational solutions to the mKP equation has been made 
in \cite{Iliev07} (see also \cite{Z14}): again, the 
evolution of poles with respect to the higher times $t_k$ 
of the mKP hierarchy is governed by 
the higher Hamiltonians $\mbox{tr}\, L^k$ of the Ruijsenaars-Schneider system. 
Recently this result was generalized to trigonometric solutions \cite{PZ19}. 
However, the corresponding result for more general elliptic solutions was 
missing in the literature.

In this paper we study the correspondence of the 
2DTL hierarchy and the Ruijsenaars-Schneider
hierarchy for elliptic solutions of the former. Our method consists in a 
solution of the auxiliary linear problems for the wave function and its adjoint using a
suitable pole ansatz. The tau-function of the 2DTL hierarchy for elliptic 
solutions has the form
\beq\label{tauell}
\tau (x, {\bf t}, \bar {\bf t})=\exp \Bigl (-\sum_{k\geq 1}kt_k \bar t_k \Bigr )
\prod_{i=1}^{N}\sigma (x-x_i({\bf t}, \bar {\bf t})),
\eeq
where $\sigma (x)$ is the Weierstrass $\sigma$-function
with quasi-periods $2\omega$, $2\omega '$ such that
${\rm Im} (\omega '/\omega )>0$ (the definition is
given below in section \ref{section:elliptic}). 
The zeros $x_i$ of the tau-function are poles of the 
solution. They are assumed to be 
all distinct.

We show that the dynamics of poles in the times ${\bf t}$, $\bar {\bf t}$ is Hamiltonian
and identify the corresponding Hamiltonians which turn out to be higher Hamiltonians of the 
elliptic Ruijesenaars-Schneider system. The generating function of the Hamiltonians is
$\lambda (z)$, where the spectral parameters $\lambda , z$ are connected by the equation
of the spectral curve
\beq\label{curve}
\det_{N\times N} \Bigl (ze^{\eta \zeta (\lambda )}I -L(\lambda )\Bigr )=0,
\eeq
where $\zeta (\lambda )$ is the Weierstrass $\zeta$-function, $I$ is the unity matrix and
$L(\lambda )$ is the Lax matrix. Any point of the spectral curve is $P=(z, \lambda )$, where
$z, \lambda$ are connected by equation (\ref{curve}).
There are to distinguished points on the spectral curve:
$P_{\infty}=(\infty , 0)$ and $P_0=(0, N\eta )$. The Hamiltonians corresponding to 
the positive time flows ${\bf t}$ are coefficients of the expansion of the function 
$\lambda (z)$ in negative powers of $z$ around the point $P_{\infty}$ while
the Hamiltonians corresponding to 
the negative time flows $\bar {\bf t}$ are coefficients of the expansion of the function 
$\lambda (z)$ in positive powers of $z$ around the point $P_{0}$. This is the main
result of the paper. 

\subsection{Elliptic Ruijsenaars-Schneider model}

Here we collect the main facts on the elliptic Ruijsenaars-Schneider system following
the paper \cite{Ruij87}. 

The $N$-particle elliptic Ruijsenaars-Schneider system is a completely integrable model.
It can be regarded as a relativistic extension of the Calogero-Moser system. 
The dynamical variables are coordinates $x_i$ and momenta $p_i$ with 
canonical Poissson brackets $\{x_i, p_j\}=\delta_{ij}$. 
The integrals of motion in involution have the form
\beq\label{intr1}
I_k= \sum_{I\subset \{1, \ldots , N\}, \, |I|=k}
\exp \Bigl (\sum_{i\in I}p_i\Bigr ) \prod_{i\in I, j\notin I}\frac{\sigma 
(x_i-x_j+\eta )}{\sigma (x_i-x_j)}, \quad k=1, \ldots , N,
\eeq
where $\eta$ is a parameter (the inverse velocity of light). 
In particular,
\beq\label{intr2}
I_1= \sum_i e^{p_i}\prod_{j\neq i} \frac{\sigma 
(x_i-x_j+\eta )}{\sigma (x_i-x_j)}, \qquad
I_N=\exp \Bigl (\sum_{i=1}^{N}p_i\Bigr ).
\eeq
Comparing to the paper \cite{Ruij87}, our formulas differ by the canonical transformation
$$
e^{p_i}\to e^{p_i}\prod_{j\neq i}\left ( \frac{\sigma 
(x_i-x_j+\eta )}{\sigma (x_i-x_j-\eta )}\right )^{1/2}, \quad x_i\to x_i,
$$
which allows one to eliminate square roots in \cite{Ruij87}. The Hamiltonian 
of the model is $H_1=I_1$. 

The velocities of the particles are
\beq\label{intr4}
\dot x_i =\frac{\p H_1}{\p p_i}=e^{p_i}\prod_{j\neq i} \frac{\sigma 
(x_i-x_j+\eta )}{\sigma (x_i-x_j)}.
\eeq
The Hamiltonian equations $\dot p_i=-\p H_1/ \p x_i$ are equivalent to the following
equations of motion:
\beq\label{te4}
\begin{array}{lll}
\ddot x_i &=&\displaystyle{-\sum_{k\neq i}\dot x_i\dot x_k \Bigl (
\zeta (x_i-x_k+\eta )+\zeta (x_i-x_k-\eta )-2\zeta (x_i-x_k)\Bigr )}
\\ && \\
&=&\displaystyle{\sum_{k\neq i}\dot x_i\dot x_k\frac{\wp '(x_i-x_k)}{\wp (\eta )-
\wp (x_i-x_k)},}
\end{array}
\eeq
where $\zeta (x)$ and $\wp (x)$ are Weierstrass $\zeta$- and $\wp$-functions. 

One can also introduce integrals of motion $I_{-k}$ as
\beq\label{intr1a}
I_{-k}=I_{N}^{-1}I_{N-k}=\sum_{I\subset \{1, \ldots , N\}, \, |I|=k}
\exp \Bigl (-\sum_{i\in I}p_i\Bigr ) \prod_{i\in I, j\notin I}\frac{\sigma 
(x_i-x_j-\eta )}{\sigma (x_i-x_j)}.
\eeq
In particular, 
\beq\label{intr2a}
I_{-1}= \sum_i e^{-p_i}\prod_{j\neq i} \frac{\sigma 
(x_i-x_j-\eta )}{\sigma (x_i-x_j)}.
\eeq
It is natural to put $I_0=1$. It can be easily verified that equations of motion 
in the time $\bar t_1$ corresponding to the Hamiltonian $\bar H_1=I_{-1}$ are the same
equations (\ref{te4}).

\subsection{Organization of the paper}

The paper is organized as follows. In section 2 we remind the reader the main facts
about the 2DTL hierarchy. We recall the Lax formulation in terms of pseudo-difference 
operators, the bilinear identity for the tau-function and auxiliary linear problems
for the wave function. 
Section 3 is devoted to solutions which are elliptic functions
of $x=t_0$. We introduce double-Bloch solutions to the auxiliary linear problem
and express them as a linear combination of elementary double-Bloch functions having
just one simple pole in the fundamental domain. 
In section 4 we obtain equations of motion for the poles as functions 
of the time $t_1$ together with their Lax representation. Properties of the spectral curve
are discussed in section 5.
In section 6 we consider the dynamics of poles with respect to the higher times and derive
the corresponding Hamiltonian equations. 
Rational and trigonometric limits are addressed
in section 7. Explicit examples of the Hamiltonians are given
in section 8.

\section{The 2D Toda latttice hierarchy}

Here we very briefly review the 2DTL hierarchy (see \cite{UT84}).
Let us consider the pseudo-difference Lax operators 
\beq\label{mkp1}
{\cal L}=e^{\eta \p_x}+\sum_{k\geq 0}U_k(x) e^{-k\eta \p_x}, \quad
\bar {\cal L}=c(x)e^{-\eta \p_x}+\sum_{k\geq 0}\bar U_k(x) e^{k\eta \p_x},
\eeq
where $e^{\eta \p_x}$ is the shift operator acting as 
$e^{\pm \eta \p_x}f(x)=f(x\pm \eta )$ and
the coefficient functions $U_k$, $\bar U_k$ 
are functions of $x$, ${\bf t}$, $\bar {\bf t}$.
The equations of the hierarchy are differential-difference
equations for the functions $c$, $U_k$, $\bar U_k$. They are encoded in the Lax equations
\beq\label{mkp2}
\p_{t_m}{\cal L}=[{\cal B}_m, {\cal L}], \quad
\p_{t_m}\bar {\cal L}=[{\cal B}_m, \bar {\cal L}]
\qquad {\cal B}_m=({\cal L}^m)_{\geq 0},
\eeq
\beq\label{mkp2a}
\p_{\bar t_m}{\cal L}=[\bar {\cal B}_m, {\cal L}], \quad
\p_{\bar t_m}\bar {\cal L}=[\bar {\cal B}_m, \bar {\cal L}]
\qquad \bar {\cal B}_m=(\bar {\cal L}^m)_{< 0},
\eeq
where $\displaystyle{\Bigl (\sum_{k\in \z} U_k e^{k\eta \p_x}\Bigr )_{\geq 0}=
\sum_{k\geq 0} U_k e^{k\eta \p_x}}$,
$\displaystyle{\Bigl (\sum_{k\in \z} U_k e^{k\eta \p_x}\Bigr )_{< 0}=
\sum_{k<0} U_k e^{k\eta \p_x}}$
For example, ${\cal B}_1=e^{\eta \p_x}+U_0(x)$, $\bar {\cal B}_1=c(x)e^{-\eta \p_x}$.

An equivalent formulation is through the zero
curvature (Zakharov-Shabat) equations
\beq\label{mkp3}
\p_{t_n}{\cal B}_m -\p_{t_m}{\cal B}_n +[{\cal B}_m, {\cal B}_n]=0,
\eeq
\beq\label{mkp3a}
\p_{\bar t_n}{\cal B}_m -\p_{t_m}\bar {\cal B}_n +[{\cal B}_m, \bar {\cal B}_n]=0,
\eeq
\beq\label{mkp3b}
\p_{\bar t_n}\bar {\cal B}_m -\p_{\bar t_m}\bar {\cal B}_n +[\bar {\cal B}_m, \bar {\cal B}_n]=0.
\eeq
For example, 
from (\ref{mkp3a}) at $m=n=1$ we have
$$
\left \{ \begin{array}{l}
\p_{t_1}\log c(x)=v(x)-v(x-\eta )
\\ \\
\p_{\bar t_1} v(x)=c(x)-c(x+\eta ),
\end{array}
\right.
$$
where $v=U_0$. 
Excluding $v(x)$, we get the second order differential-difference equation for 
$c(x)$:
$$
\p_{t_1}\p_{\bar t_1}\log c(x)=2c(x)-c(x+\eta )-c(x-\eta )
$$
which is one of the forms of the 2D Toda equation. After the change of variables
$c(x)=e^{\varphi (x)-\varphi (x-\eta )}$ it acquires the most familiar form
\beq\label{mkp5a}
\p_{t_1}\p_{\bar t_1}\varphi (x)=e^{\varphi (x)-\varphi (x-\eta )}-
e^{\varphi (x+\eta )-\varphi (x)}.
\eeq

The zero curvature equations
are compatibility conditions for the auxiliary linear problems
\beq\label{mkp6}
\p_{t_m}\psi ={\cal B}_m (x)\psi , \quad
\p_{\bar t_m}\psi =\bar {\cal B}_m (x)\psi ,
\eeq
where the wave function $\psi$ depends on a spectral parameter $z$: 
$\psi =\psi (z; {\bf t})$. The wave function has the following expansion in powers of
$z$:
\beq\label{mkp7}
\psi =z^{x/\eta}
e^{\xi ({\bf t}, z)}\left (1+ \frac{\xi_1(x, {\bf t}, \bar {\bf t})}{z}+
\frac{\xi_2(x, {\bf t}, \bar {\bf t})}{z^2}+\ldots \right ),
\eeq
where
\beq\label{mkp8}
\xi ({\bf t}, z)=\sum_{k\geq 1}t_k z^k.
\eeq
The wave operator is the pseudo-difference operator of the form
\beq\label{mkp9}
{\cal W}(x)=1+\xi_1(x)e^{-\eta \p_x}+\xi_2(x)e^{-2\eta \p_x}+\ldots \, 
\eeq
with the same coefficient functions $\xi_k$ as in 
(\ref{mkp7}), then the wave function can be written as
\beq\label{mkp10}
\psi = {\cal W}(x)z^{x/\eta} e^{\xi ({\bf t}, z)}.
\eeq
The adjoint wave function $\psi^{\dag}$ is defined by the formula
\beq\label{mkp11}
\psi^{\dag}=({\cal W}^{\dag}(x\! -\! \eta ))^{-1}z^{-x/\eta} e^{-\xi ({\bf t}, z)}
\eeq
(see, e.g., \cite{Z18}), 
where the adjoint difference operator is defined according to the rule
$(f(x) \circ e^{n\eta \p_x})^{\dag}=e^{-n\eta \p_x}\circ f(x)$. The auxiliary linear
problems for the adjoint wave function have the form
\beq\label{mkp12}
-\p_{t_m}\psi ^{\dag}={\cal B}_{m}^{\dag}(x\! -\! \eta )\psi^{\dag}.
\eeq
In particular, we have:
\beq\label{mkp13}
\begin{array}{l}
\phantom{-}\p_{t_1}\psi (x)=\psi (x+\eta )+v(x)\psi (x), 
\\ \\
-\p_{t_1}\psi^{\dag} (x)=\psi^{\dag} (x-\eta )+v(x-\eta )\psi^{\dag} (x),
\end{array}
\eeq
\beq\label{mkp13a}
\p_{\bar t_1}\psi (x)=c(x)\psi (x-\eta ).
\eeq

A common solution to the 2DTL hierarchy is provided by the tau-function $\tau =
\tau (x, {\bf t}, \bar {\bf t})$ \cite{DJKM83,JM83}. 
The tau-function satisfies the bilinear relation
\beq\label{mkp14b}
\begin{array}{c}
\displaystyle{\oint_{\infty}z^{\frac{x-x'}{\eta}-1}e^{\xi ({\bf t}, z)-\xi ({\bf t}', z)}
\tau \Bigl (x, {\bf t}-[z^{-1}], \bar {\bf t}\Bigr )
\tau \Bigl (x'+\eta , {\bf t}'+[z^{-1}], \bar {\bf t}'\Bigr )dz}
\\ \\
\displaystyle{=\, \oint_{0}z^{\frac{x-x'}{\eta}-1}
e^{\xi (\bar {\bf t}, z^{-1})-\xi (\bar {\bf t}', z^{-1})}
\tau \Bigl (x+\eta , {\bf t}, \bar {\bf t}-[z]\Bigr )
\tau \Bigl (x' , {\bf t}', \bar {\bf t}'+[z]\Bigr )dz
}
\end{array}
\eeq
valid for all $x, x'$, ${\bf t}, {\bf t}'$, $\bar {\bf t}, \bar {\bf t}'$, where
\beq\label{mkp15}
{\bf t}\pm [z]=\Bigl \{t_1\pm z, t_2\pm \frac{1}{2}z^{2}, 
t_3\pm \frac{1}{3}z^{3}, \ldots \Bigl \}.
\eeq
The integration contour in the left hand side 
is a big circle around infinity separating the singularities
coming from the exponential factor from those coming from the tau-functions. 
The integration contour in the right hand side 
is a small circle around zero separating the singularities
coming from the exponential factor from those coming from the tau-functions.

The coefficient functions of the Lax operators can be expressed through the 
tau-function. In particular,
\beq\label{mkp17}
U_0(x)=v(x)=\p_{t_1}\log \frac{\tau (x+\eta )}{\tau (x)}, \quad
c(x)=\frac{\tau (x+\eta )\tau (x-\eta )}{\tau^2(x)}.
\eeq
The Toda equation (\ref{mkp5a}) in terms of the tau-function becomes
\beq\label{mkp18a}
\p_{t_1}\p_{\bar t_1}\log \tau (x)=-\frac{\tau (x+\eta )\tau (x-\eta )}{\tau^2(x)}.
\eeq

The wave function and its adjoint are expressed through the tau-function according to
the formulas
\beq\label{mkp19}
\psi =z^{x/\eta}e^{\xi ({\bf t}, z)}\frac{\tau (x, {\bf t}-[z^{-1}], 
\bar {\bf t})}{\tau (x, {\bf t}, \bar {\bf t})},
\eeq
\beq\label{mkp20}
\psi^{\dag} =z^{-x/\eta}e^{-\xi ({\bf t}, z)}
\frac{\tau (x, {\bf t}+[z^{-1}], \bar {\bf t})}{\tau (x, {\bf t}, \bar {\bf t})}.
\eeq
One may also introduce the complimentary wave functions $\bar \psi$, $\bar \psi^{\dag}$ 
by the formulas
\beq\label{mkp19a}
\bar \psi =z^{x/\eta}e^{\xi (\bar {\bf t}, z^{-1})}\frac{\tau (x+\eta , {\bf t}, 
\bar {\bf t}-[z])}{\tau (x, {\bf t}, \bar {\bf t})},
\eeq
\beq\label{mkp20a}
\bar \psi^{\dag} =z^{-x/\eta}e^{-\xi (\bar {\bf t}, z^{-1})}
\frac{\tau (x-\eta , {\bf t}, \bar {\bf t}+[z])}{\tau (x, {\bf t}, \bar {\bf t})}.
\eeq
They satisfy the same auxiliary linear problems as the wave functions $\psi$, $\psi^{\dag}$.
It will be more convenient for us to work with the renormalized wave functions
\beq\label{mkp19b}
\phi (x)=\frac{\tau (x)}{\tau (x+\eta )}\, \bar \psi (x)=
z^{x/\eta}e^{\xi (\bar {\bf t}, z^{-1})}\frac{\tau (x+\eta , {\bf t}, 
\bar {\bf t}-[z])}{\tau (x+\eta , {\bf t}, \bar {\bf t})},
\eeq
\beq\label{mkp20b}
\phi^{\dag}(x)=\frac{\tau (x)}{\tau (x-\eta )}\, \bar \psi^{\dag}(x)=
z^{-x/\eta}e^{-\xi (\bar {\bf t}, z^{-1})}
\frac{\tau (x-\eta , {\bf t}, \bar {\bf t}+[z])}{\tau (x-\eta , {\bf t}, \bar {\bf t})}.
\eeq
They satisfy the linear equations
\beq\label{mkp20c}
\p_{\bar t_1}\phi (x)=\phi (x-\eta )-\bar v(x)\phi (x), \quad
-\p_{\bar t_1}\phi^{\dag} (x)=\phi^{\dag} (x+\eta )-\bar v(x-\eta )\phi ^{\dag} (x),
\eeq
where $\displaystyle{\bar v(x)=\p_{\bar t_1}\log \frac{\tau (x+\eta )}{\tau (x)}}$.

Finally, let us point out useful corollaries of the bilinear relation (\ref{mkp14b}).
Differentiating it with respect to $t_m$ and putting $x=x'$, ${\bf t}={\bf t}'$ 
$\bar {\bf t}=\bar {\bf t}'$ after that, we obtain:
\beq\label{mkp21}
\begin{array}{c}
\displaystyle{
\frac{1}{2\pi i}
\oint_{\infty}z^{m-1}
\tau \Bigl (x, {\bf t}-[z^{-1}]\Bigr )\tau \Bigl (x+\eta , {\bf t}+[z^{-1}]\Bigr )dz}
\\ \\
\displaystyle{ =
\p_{t_m}\tau (x+\eta , {\bf t})\tau (x, {\bf t})-
\p_{t_m}\tau (x, {\bf t})\tau (x+\eta , {\bf t})}
\end{array}
\eeq
or
\beq\label{mkp22}
\res_{\infty}\, \Bigl (z^m \psi (x)\psi^{\dag}(x+\eta )\Bigr )=\p_{t_m}
\log \frac{\tau (x+\eta )}{\tau (x)}.
\eeq
Equivalently, equation (\ref{mkp22}) can be written in the form
\beq\label{mkp22b}
\psi (x)\psi^{\dag}(x+\eta )=1+\sum_{m\geq 1}z^{-m-1}\p_{t_m}
\log \frac{\tau (x+\eta )}{\tau (x)}.
\eeq
In a similar way, differentiating the bilinear relation (\ref{mkp14b}) 
with respect to $\bar t_m$ and putting $x=x'$, ${\bf t}={\bf t}'$, 
$\bar {\bf t}=\bar {\bf t}'$ after that,
we obtain the relation
\beq\label{mkp22a}
\res_{0}\, \Bigl (z^{-m} \phi (x)\phi^{\dag}(x+\eta )\Bigr )=-\p_{\bar t_m}
\log \frac{\tau (x+\eta )}{\tau (x)}.
\eeq
Here $\displaystyle{\res_{\infty}}$, $\displaystyle{\res_{0}}$ are defined according to the convention $\displaystyle{\res_{\infty}}(z^{-n})=\delta_{n1}$, 
$\displaystyle{\res_{0}}(z^{-n})=\delta_{n1}$.

\section{Elliptic solutions to the Toda equation and double-Bloch wave functions}

\label{section:elliptic}

The ansatz for the tau-function of elliptic (double-periodic in the complex plane)
solutions to the 2DTL hierarchy is given by (\ref{tauell}). 
In (\ref{tauell})
$$
\sigma (x)=\sigma (x |\, \omega , \omega ')=
x\prod_{s\neq 0}\Bigl (1-\frac{x}{s}\Bigr )\, e^{\frac{x}{s}+\frac{x^2}{2s^2}},
\quad s=2\omega m+2\omega ' m' \quad \mbox{with integer $m, m'$}
$$ 
is the Weierstrass 
$\sigma$-function with quasi-periods $2\omega$, $2\omega '$ such that 
${\rm Im} (\omega '/ \omega )>0$. 
It is connected with the Weierstrass 
$\zeta$- and $\wp$-functions by the formulas $\zeta (x)=\sigma '(x)/\sigma (x)$,
$\wp (x)=-\zeta '(x)=-\p_x^2\log \sigma (x)$.
The monodromy properties of the function $\sigma (x)$ 
are
\beq\label{e1a}
\sigma (x+2\omega )=-e^{2\zeta (\omega ) (x+\omega )}\sigma (x), \quad
\sigma (x+2\omega ' )=-e^{2\zeta (\omega ' ) (x+\omega ' )}\sigma (x),
\eeq
where the constants $\zeta (\omega )$, $\zeta (\omega ' )$ are  related by
$\zeta (\omega )\omega ' -\zeta (\omega ') \omega =\pi i /2$.

For elliptic solutions
\beq\label{e2}
v(x)=\sum_{i}\dot x_i \Bigl (\zeta (x-x_i)-\zeta (x-x_i+\eta )\Bigr )
\eeq
is an elliptic function,
so one can find {\it double-Bloch} solutions to the linear problem (\ref{mkp15}). 
The double-Bloch function satisfies the monodromy properties
$\psi (x+2\omega )=B\psi (x)$, $\psi (x+2\omega ' )=B'\psi (x)$ with some
Bloch multipliers $B$, $B'$. Any non-trivial double-Bloch function (i.e. not an exponential
function) must have poles in $x$ in the fundamental domain. 
The Bloch multipliers of the wave function
(\ref{mkp19}) are:
\beq\label{e3}
B=z^{2\omega /\eta}\exp \Bigl (-2\zeta (\omega )\sum_i (e^{-D(z)}\! -\! 1)x_i )\Bigr ),
\quad
B'=z^{2\omega ' /\eta}\exp \Bigl (-2\zeta (\omega ' )\sum_i (e^{-D(z)}\! -\! 1)x_i)\Bigr ),
\eeq
where the differential operator $D(z)$ is
\beq\label{D(z)}
D(z)=\sum_{k\geq 1}\frac{z^{-k}}{k}\, \p_{t_k}.
\eeq
It follows from the equation (\ref{mkp22b}) that the Bloch multipliers of the 
function $\psi^{\dag}$ are $1/B$, $1/B'$. Indeed, the right hand side of 
(\ref{mkp22b}) is an elliptic function of $x$, therefore, the left hand side 
must be also an elliptic function. 

Let us introduce the 
elementary double-Bloch function $\Phi (x, \lambda )$ defined as
\beq\label{Phi}
\Phi (x, \lambda )=\frac{\sigma (x+\lambda )}{\sigma (\lambda )\sigma (x)}\,
e^{-\zeta (\lambda )x}
\eeq
($\zeta (\lambda )$ is the Weierstrass $\zeta$-function).
The monodromy properties of the function $\Phi$ are
$$
\Phi (x+2\omega , \lambda )=e^{2(\zeta (\omega )\lambda - \zeta (\lambda )\omega )}
\Phi (x, \lambda ),
$$
$$
\Phi (x+2\omega ' , \lambda )=e^{2(\zeta (\omega ' )\lambda - \zeta (\lambda )\omega ' )}
\Phi (x, \lambda ),
$$
so it is indeed a double-Bloch function.
The function $\Phi$
has a simple pole
at $x=0$ with residue 1: 
$$
\Phi (x, \lambda )=\frac{1}{x}+\alpha_1 x  +\ldots , \qquad 
x\to 0,
$$
where $\alpha_1=-\frac{1}{2}\, \wp (\lambda )$. 
When this does not lead to a misunderstanding, 
we will suppress the second argument of $\Phi$ writing simply 
$\Phi (x)=\Phi (x, \lambda )$. 
We will also need the $x$-derivative 
$\Phi '(x, \lambda )=\p_x \Phi (x, \lambda )$.

Below we need the following identities satisfied by the function $\Phi$:
\beq\label{id0}
\Phi (x, -\lambda )=-\Phi (-x, \lambda ),
\eeq
\beq\label{id1}
\Phi '(x)=\Phi (x)(\zeta (x+\lambda )-\zeta (\lambda )-\zeta (x)),
\eeq
\beq\label{id2}
\Phi (x)\Phi (y)=\Phi (x+y)(\zeta (x)+\zeta (y)+\zeta (\lambda )-\zeta (x+y+\lambda )).
\eeq

Equations (\ref{mkp19}), (\ref{mkp20}) imply that the wave functions
$\psi$, $\psi^\dag$ have simple poles at the points $x_i$. One can expand the wave functions
using the elementary double-Bloch functions as follows: 
\beq\label{e5}
\psi =k^{x/\eta}e^{\xi({\bf t}, z)}\sum_i c_i \Phi (x-x_i, \lambda )
\eeq
\beq\label{e5a}
\psi^\dag =k^{-x/\eta}e^{-\xi({\bf t}, z)}\sum_i c^*_i \Phi (x-x_i, -\lambda )
\eeq
(this is similar to expansion of a rational function
in a linear combination of simple fractions). Here $c_i$, $c^*_i$
are expansion coefficients which do not depend on $x$ and 
$k$ is an additional spectral parameter. Note that the normalization of the functions
(\ref{mkp19}), (\ref{mkp20}) implies 
that $c_i$ and $c_i^*$ are $O(\lambda )$ as $\lambda \to 0$.
One can see that (\ref{e5}) is 
a double-Bloch function with Bloch multipliers
\beq\label{e6}
B=e^{\frac{2\omega}{\eta} (\log k-\eta \zeta (\lambda )) + 2\zeta (\omega )\lambda }, \qquad
B '=e^{\frac{2\omega '}{\eta} (\log k-\eta \zeta (\lambda )) + 2\zeta (\omega ' )\lambda }
\eeq
and (\ref{e5a}) has Bloch multipliers $B^{-1}$ and $B^{' -1}$. These Bloch multipliers
should coincide with (\ref{e3}). 

Therefore, comparing (\ref{e3}) with (\ref{e6}), we get
$$
\frac{2\omega}{\eta} \Bigl (\log (k/z)-\eta \zeta (\lambda )\Bigr )+
2\zeta (\omega )\Bigl (\lambda +(e^{-D(z)}-1)\sum_i x_i\Bigr )=
2\pi i n,
$$
$$
\frac{2\omega '}{\eta} \Bigl (\log (k/z)-\eta \zeta (\lambda )\Bigr )+
2\zeta (\omega ' )\Bigl (\lambda +(e^{-D(z)}-1)\sum_i x_i\Bigr )=
2\pi i n'
$$
with some integer $n, n'$. Regarding these equations as a linear system, we obtain the solution
$$
\log (k/z)-\eta \zeta (\lambda)=2n'\eta \zeta (\omega)-2n\eta \zeta (\omega '),
$$
$$
\lambda +(e^{-D(z)}-1)\sum_i x_i=2n\omega ' -2n'\omega .
$$
Shifting $\lambda$ by a suitable vector of the lattice spanned by $2\omega$, $2\omega '$,
one gets zeros in the right hand sides of these equalities, so we can 
represent the connection between the spectral parameters $k,z, \lambda$ in the form
\beq\label{e7}
\left \{\begin{array}{l}
k=z\, e^{\eta\zeta (\lambda )},
\\ \\
\displaystyle{\lambda = (1-e^{-D(z)}) \sum_i x_i}.
\end{array} \right.
\eeq
These two equations for three spectral parameters $k, z, \lambda$ determine the spectral curve. 
Another description of the same spectral curve is obtained below as 
the spectral curve of the Ruijsenaars-Schneider system 
(it is given by the characteristic polynomial of the Lax matrix
$L(\lambda )$ for the Ruijsenaars-Schneider system). 
It appears in the form $R(k,\lambda)=0$, where
$R(k,\lambda)$ is a polynomial in $k$ whose coefficients are elliptic functions of
$\lambda$ (see below section \ref{section:curve}). 
These coefficients are integrals of motion in involution. The spectral curve
in the form $R(k,\lambda)=0$ appears if one excludes $z$ from the equations (\ref{e7}).
Equivalently, one can represent the spectral curve as a relation connecting 
$z$ and $\lambda$:
\beq\label{e7a}
R(ze^{\eta \zeta (\lambda )}, \lambda )=0.
\eeq

The second equation in (\ref{e7}) can be written as expansion in powers of $z$:
\beq\label{h2}
\lambda =\lambda (z)= -\sum_{m\geq 1}z^{-m}\hat h_m {\cal X}, \qquad {\cal X}:=\sum_i x_i.
\eeq
Here $\hat h_k$ are differential operators of the form
\beq\label{h1}
\hat h_m = -\frac{1}{m}\, \p_{t_m}+\,\,\, \mbox{higher order operators
in $\p_{t_1}, \p_{t_2}, \ldots , \p_{t_{m-1}}$}.
\eeq
The first few are
$$
\hat h_1 =-\p_{t_1}, \quad 
\hat h_2 =\frac{1}{2}\, (\p_{t_1}^2 -\p_{t_2}),
\hat h_3 =\frac{1}{6}\, (-\p_{t_1}^3 +3\p_{t_1}\p_{t_2}-2\p_{t_3}).
$$
As is mentioned above, the coefficients in the expansion (\ref{h2}) are integrals of
motion, i.e., $\p_{t_j}\hat h_m {\cal X}=0$ for all $j,m$. It then follows from 
the explicit form of the operators $\hat h_m$ that
$\p_{t_j}\p_{t_1}{\cal X}=0$. A simple inductive argument then shows that
$\p_{t_j}\p_{t_m}{\cal X}=0$ for all $j,m$. This means that
$-\hat h_m {\cal X}=\frac{1}{m}\, \p_{t_m}{\cal X}$ and so ${\cal X}$ is a linear function of times:
\beq\label{h3}
{\cal X}=\sum_i x_i ={\cal X}_0 +\sum_{m\geq 1}V_m t_m
\eeq
with some constants $V_m$ (which are velocities of the ``center of masses'' of the points $x_i$
with respect to the times $t_m$ multiplied by $N$). 
Therefore, the second equation in (\ref{e7}) can be written as
\beq\label{e8}
\lambda = D(z)\sum_i x_i=\sum_{j\geq 1}\frac{z^{-j}}{j}\, V_j.
\eeq
In what follows we will show that $H_m=\frac{1}{m}\, V_{m}$ are Hamiltonians
for the dynamics of the poles in $t_m$, with $H_1$ being the standard Ruijsenaars-Schneider
Hamiltonian. 

\section{Dynamics of poles with respect to $t_1$}

The next procedure is standard after the work \cite{Krichever80}. We substitute
$v(x)$ in the form (\ref{e2}) and $\psi$ in the form (\ref{e5}) into the left hand side
of the linear problem
$$
\p_{t_1}\psi (x) -\psi (x+\eta )-v(x)\psi (x)=0
$$
and cancel the poles at the points $x=x_i$, $x=x_i-\eta$. 
The highest poles are of second order but it is easy to see that they cancel identically.
A simple calculation shows that
cancellation of first order poles leads to the conditions
$$
\left \{ \begin{array}{l}
\displaystyle{zc_i+\dot c_i= \dot x_i \sum_{j\neq i}c_k \Phi (x-x_j)
+c_i\sum_{j\neq i}\dot x_j \zeta (x_i-x_j)-
c_i\sum_{j}\dot x_j\zeta (x_i-x_j+\eta )}
\\ \\
\displaystyle{kc_i-\dot x_i \sum_j c_j \Phi (x_i-x_j-\eta )=0}.
\end{array}
\right.
$$
Here and below dot means the $t_1$-derivative.
These conditions can be written in the matrix form 
as a system of linear equations for the vector
${\bf c}=(c_1, \ldots , c_N)^T$:
\beq\label{ell9}
\left \{ \begin{array}{l}
L(\lambda){\bf c}=k{\bf c}
\\ \\
\dot {\bf c}=M(\lambda) {\bf c},
\end{array} \right.
\eeq
where $N\! \times \! N$ matrices $L$, $M$ are
\beq\label{ell7}
L_{ij}=\dot x_i\Phi (x_i-x_j-\eta ),
\eeq
\beq\label{ell8}
\begin{array}{l}
\displaystyle{M_{ij}=\delta_{ij}\Bigl (k-z +\sum_{l\neq i}\dot x_l \zeta (x_i-x_l)-
\sum_{l}\dot x_l \zeta (x_i-x_l+\eta )\Bigr )}
\\ \\
\phantom{aaaaaaaaaaaaaaaaaaaa}+(1-\delta_{ij})
\dot x_i\Phi (x_i-x_j)-\dot x_i\Phi (x_i-x_j-\eta).
\end{array}
\eeq
They depend on the spectral parameter $\lambda$. 

Differentiating the first equation in (\ref{ell9}) with respect to $t_1$, 
and substituting the second equation, we obtain
the compatibility condition of the linear problems (\ref{ell9}):
\beq\label{ell12}
\Bigl (\dot L+[L,M]\Bigr ) {\bf c} =0.
\eeq
The Lax equation $\dot L+[L,M]=0$ is equivalent to the equations of motion 
of the elliptic Ruijsenaars-Schneider system (see
\cite{Z19} for details of the calculation). Note that
our matrix $M$ differs from the standard one by the term $\delta_{ij}(k\! -\! z)$ but
it does not affect the compatibility condition because it is proportional to the unity matrix.
It follows from the Lax representation that the time evolution is an isospectral
transformation of the Lax matrix $L$, so all traces $\mbox{tr}\, L^m$ and the 
characteristic polynomial $\det (kI-L)$, where $I$ is the unity matrix, 
are integrals of motion.
Note that the Lax matrix is written in terms of the momenta $p_i$ as follows:
\beq\label{e9}
L_{ij}=\Phi (x_i-x_j-\eta )e^{p_i}\prod_{l\neq i}\frac{\sigma (x_i-x_l+\eta )}{\sigma (x_i-x_l)}.
\eeq

A similar calculation shows that the adjoint linear problem for the function (\ref{e5a})
leads to the equations
\beq\label{ell9a}
\left \{ \begin{array}{l}
{\bf c}^{*T}\dot X^{-1}L(\lambda)\dot X =k{\bf c}^{*T}
\\ \\
\dot {\bf c}^{*T}=-{\bf c}^{*T}M^{*}(\lambda) ,
\end{array} \right.
\eeq
where $\dot X =\mbox{diag}\, (\dot x_1, \ldots , \dot x_N)$ and
\beq\label{ell8a}
\begin{array}{l}
\displaystyle{M^{*}_{ij}=\delta_{ij}\Bigl (k-z -\sum_{l\neq i}\dot x_l \zeta (x_i-x_l)+
\sum_{l}\dot x_l \zeta (x_i-x_l-\eta )\Bigr )}
\\ \\
\phantom{aaaaaaaaaaaaaaaaaaaa}+(1-\delta_{ij})
\dot x_j\Phi (x_i-x_j)-\dot x_j\Phi (x_i-x_j-\eta).
\end{array}
\eeq

\section{The spectral curve}

\label{section:curve}

The first of the equations (\ref{ell9}) determines a connection between
the spectral parameters $k, \lambda$ which is the equation of the spectral curve:
\beq\label{spec1}
R(k, \lambda )=\det \Bigl (kI-L(\lambda )\Bigr )=0.
\eeq
As it was already mentioned, the spectral curve 
is an integral of motion.
The matrix
$L=L(\lambda )$, which has an essential singularity at $\lambda =0$, can be 
represented in the form $L(\lambda )=e^{\eta \zeta (\lambda )}
V\tilde L(\lambda ) V^{-1}$, where matrix elements of 
$\tilde L(\lambda )$ do not have 
essential singularities and $V$ is the diagonal matrix $V_{ij}=\delta_{ij}
e^{-\zeta (\lambda )x_i}$. The matrix $\tilde L(\lambda )$ reads
\beq\label{spec2}
\tilde L_{ij}(\lambda )=\frac{\sigma (x_i-x_j-\eta +\lambda )}{\sigma (\lambda )
\sigma (x_i-x_j-\eta )}\, e^{p_i}
\prod_{l\neq i}\frac{\sigma (x_i-x_l+\eta )}{\sigma (x_i-x_l)}.
\eeq
Using the connection (\ref{e7}) between $k$ and $z$, we represent the spectral curve
in the form
\beq\label{spec3}
\det \Bigl (zI-\tilde L(\lambda )\Bigr )=0.
\eeq
The spectral curve is an $N$-sheet covering over the $\lambda$-plane. 
Any point of the curve is $P=(z, \lambda )$, where $z, \lambda$ are connected by equation
(\ref{spec3}) and there are $N$ points above each point $\lambda$. 

In order to represent the spectral curve in explicit form, we recall the
identity
\beq\label{spec4}
\det_{1\leq i,j\leq N}\left (
\frac{\sigma (x_i-y_j +\lambda )}{\sigma (\lambda )\sigma (x_i-y_j)}\right )
=\frac{\sigma \Bigl (\lambda +\sum\limits_{i=1}^N (x_i-y_i)\Bigr )}{\sigma (\lambda )}
\frac{\prod\limits_{i<j}\sigma (x_i-x_j)\sigma (y_j-y_i)}{\prod\limits_{i,j}\sigma (x_i-y_j)}
\eeq
for the determinant of the elliptic Cauchy matrix. Using this identity, one can represent
(\ref{spec3}) as
\beq\label{spec5}
\sum_{n=0}^N \varphi_n(\lambda )I_n z^{N-n}=0,
\eeq
where $I_n$ are the Ruijsenaars-Schneider integrals of motion (\ref{intr1}) and
\beq\label{spec6}
\varphi _n(\lambda )=\frac{\sigma (\lambda -n\eta )}{\sigma (\lambda )
\sigma^n(\eta)}.
\eeq

Let us fix two distinguished points on the spectral curve. As $\lambda \to 0$, we have
$$
\tilde L(\lambda )=\dot X E \lambda^{-1} +O(1),
$$
where $E$ is the rank 1 matrix with matrix elements $E_{ij}=1$ for all $i,j$. Therefore,
using the formula for determinant of the matrix $I+Y$, where $Y$ is a rank 1 matrix, 
we can write
$$
\det \Bigl (zI-\tilde L(\lambda )\Bigr )=z^N -z^{N-1}\lambda^{-1}I_1 +O(z^{N-1}),
$$
or
$$
\det \Bigl (zI-\tilde L(\lambda )\Bigr )=(z-h_N(\lambda )\lambda^{-1})
\prod_{j=1}^{N-1}(z-h_j(\lambda )),
$$
where the functions $h_j$ are regular functions at $\lambda =0$ and $h_N(0)=I_1=H_1$. 
We see that among $N$ points above $\lambda =0$ one is distinguished: it is the point
$P_{\infty}=(\infty , 0)$. Another distinguished point is $P_0=(0, N\eta )$. Since 
$\varphi_N(N\eta )=0$, we see that this point indeed belongs to the curve. 

\section{Dynamics in higher times}

\subsection{Positive times}

In order to study the dynamics of poles in the higher positive times, 
we take advantage of the relation (\ref{mkp22}), which, after the substitution of the
wave functions for elliptic solutions acquires the form
\beq\label{pt1}
\sum_{i,j}\res_{\infty}\Bigl (z^{m-1}c_ic_j^* \Phi (x-x_i, \lambda )
\Phi (x-x_j+\eta , -\lambda )\Bigr )=
\sum_n \p_{t_m}x_n \Bigl (\zeta (x-x_n)-\zeta (x-x_n +\eta )\Bigr ).
\eeq
Equating the residues at $x=x_i-\eta$, we obtain:
\beq\label{pt2}
\p_{t_m}x_i=-\sum_j \res_{\infty}\Bigl (z^{m}k^{-1}c_i^*c_j \Phi (x_i-x_j-\eta , \lambda )\Bigr ).
\eeq
In the matrix form, we can write this relation as
\beq\label{pt3}
\p_{t_m}x_i=-\res_{\infty}\Bigl (z^{m}k^{-1}{\bf c}^{*T}\dot X^{-1}(\p_{p_i}L){\bf c}\Bigr ).
\eeq
Summing (\ref{pt3}) over $i$ and using (\ref{e8}), we can write
\beq\label{pt4}
\res_{\infty}\Bigl (z^m \lambda '(z)\Bigr )=
\res_{\infty}\Bigl (z^{m}k^{-1}{\bf c}^{*T}\dot X^{-1}L{\bf c}\Bigr ),
\eeq
where we took into account that $\sum_i \p_{p_i}L=L$. Using the fact that
${\bf c}=O(1/z)$, we conclude that
$$
k^{-1}{\bf c}^{*T}\dot X^{-1}L{\bf c}=\lambda '(z),
$$
or
\beq\label{pt5}
{\bf c}^{*T}\dot X^{-1}{\bf c}=\lambda '(z).
\eeq
Next, having (\ref{pt5}) at hand, 
we can continue (\ref{pt3}) as the following chain of equalities:
$$
\p_{t_m}x_i=-\res_{\infty}\Bigl (z^{m}k^{-1}{\bf c}^{*T}\dot X^{-1}(\p_{p_i}L){\bf c}\Bigr )
$$
$$\hspace{-2cm}
=-\res_{\infty}\Bigl (z^{m}\p_{p_i}\Bigl (k^{-1}{\bf c}^{*T}\dot X^{-1}L{\bf c}\Bigr )\Bigr )+
\res_{\infty}\Bigl (z^{m}k^{-1}\p_{p_i} ({\bf c}^{*T}\dot X^{-1})L{\bf c}\Bigr )
$$
$$\hspace{2cm}
+\res_{\infty}\Bigl (z^{m}k^{-1}{\bf c}^{*T}\dot X^{-1}L\p_{p_i}{\bf c}\Bigr )+
\res_{\infty}\Bigl (z^{m}\p_{p_i}(k^{-1}) {\bf c}^{*T}\dot X^{-1}L{\bf c}\Bigr )
$$
$$\hspace{-2cm}
=-\res_{\infty}\Bigl (z^{m}\p_{p_i}\lambda '(z)\Bigr )
+\res_{\infty}\Bigl (z^{m}\p_{p_i} ({\bf c}^{*T}\dot X^{-1}){\bf c}\Bigr )
$$
$$\hspace{2cm}
+\res_{\infty}\Bigl (z^{m}{\bf c}^{*T}\dot X^{-1}\p_{p_i}{\bf c}\Bigr )
-\res_{\infty}\Bigl (z^{m}\p_{p_i}\! \log k \, {\bf c}^{*T}\dot X^{-1}{\bf c}\Bigr )
$$
$$
=-\res_{\infty}\Bigl (z^{m}\lambda '(z)\p_{p_i}\log k\Bigr ).
$$
Here $\p_{p_i}\log k =\p_{p_i}\log k (\lambda , {\bf I})\Bigr |_{\lambda ={\rm const}}$,
where ${\bf I}$ is the full set of integrals of motion. From (\ref{e8}) we see that
$\p_{p_i}\log k = \p_{p_i}\log z (\lambda , {\bf I})\Bigr |_{\lambda ={\rm const}}$.
We consider $z$ as an independent variable, whence
$$
0=\frac{d \log z}{d p_i}=\p_{p_i}\log z\Bigr |_{\lambda ={\rm const}}
+\p_{\lambda}\log z\Bigr |_{{\bf I} ={\rm const}}\p_{p_i}\lambda
$$
or
\beq\label{pt6}
\p_{p_i}\log z=-\frac{\p_{p_i}\lambda}{z\lambda '(z)}.
\eeq
Therefore,
\beq\label{pt7}
\p_{t_m}x_i=\res_{\infty}\Bigl (z^{m-1}\p_{p_i}\lambda (z)\Bigr )=\frac{\p H_m}{\p p_i},
\eeq
where the Hamiltonian $H_m$ is given by
\beq\label{pt8}
H_m=\res_{\infty}\Bigl (z^{m-1}\lambda (z)\Bigr ).
\eeq
This is the first half of the Hamiltonian equations. 

The calculation leading to the second half of the Hamiltonian equations is rather
involved. Let us present the main steps of it. 
First of all we note that
\beq\label{pt9}
\dot X^{-1}\p_{x_i}L=[E_i, B^-] +(\tilde D^+ \! -\! \tilde D^0)E_i A^-   \!-\!
\sum_l E_l \zeta (x_l-x_i +\eta )A^- \! +\! \sum_{l\neq i}E_l \zeta (x_l-x_i )A^- :=Y_i,
\eeq
where we have introduced matrices $A^-$, $B^-$ 
and diagonal matrices $E_i$, $\tilde D^0$, $\tilde D^{\pm}$ 
defined by their matrix elements as follows:
$$
\begin{array}{l}
A^-_{jk}=\Phi (x_j-x_k-\eta ), \quad
B^-_{jk}=\Phi ' (x_j-x_k-\eta ),
\\ \\
(E_i)_{jk}=\delta_{ij}\delta_{ik}, \quad
\displaystyle{\tilde D^0_{jk}=\delta_{jk}\sum_{l\neq j}\zeta (x_j -x_l)},
\quad
\displaystyle{\tilde D^{\pm}_{jk}=\delta_{jk}\sum_{l}\zeta (x_j -x_l\pm \eta )}.
\end{array}
$$
Now, in order to prove the second half of the Hamiltonian equations it is enough to obtain
the equations
\beq\label{pt11}
\p_{t_m}p_i=\res_{\infty} \Bigl (z^{m}k^{-1}{\bf c}^{*T}Y_i{\bf c}\Bigr ),
\eeq
where $Y_i$ is the right hand side of (\ref{pt9}). Indeed, repeating the arguments after equation
(\ref{pt5}) with the change $\p_{p_i}\to \p_{x_i}$, we obtain the Hamiltonian equations
\beq\label{pt10}
\p_{t_m}p_i=-\frac{\p H_m}{\p x_i}
\eeq
with the same Hamiltonian given by (\ref{pt8}). 

It is clear that
\beq\label{pt12}
p_i=\log \dot x_i -\sum_{k\neq i}\log \frac{\sigma (x_i-x_k+\eta )}{\sigma (x_i-x_k)}.
\eeq
The $t_m$-derivative of this equality yields
\beq\label{pt12a}
\p_{t_m}p_i=\dot x_i^{-1}\p_{t_m}\dot x_i -\sum_{k\neq i}
(\p_{t_m}x_i-\p_{t_m}x_k)(\zeta (x_i-x_k+\eta )-\zeta (x_i-x_k)).
\eeq
Let us find the first term in the right hand side. For this we differentiate equation
(\ref{pt3}) which we write here in the form
$$
\p_{t_m}x_i =\res_{\infty}\Bigl (z^{m}k^{-1}{\bf c}^{*T}E_i A^{-}{\bf c}\Bigr ),
$$
with respect to $t_1$ and use the equations (\ref{ell9}), (\ref{ell9a}). In this way we get
\beq\label{pt13}
\p_{t_m}\dot x_i =\res_{\infty}\Bigl (z^{m}k^{-1}{\bf c}^{*T}
(M^* E_i A^- -E_i A^- M -E_i \dot A^-){\bf c}\Bigr ).
\eeq
Here the matrices $M$, $M^*$ are given by
\beq\label{pt14}
\begin{array}{l}
M=(k-z)I +D^0 -D^+ +\dot X A -\dot X A^-,
\\ \\
M^*=(k-z)I -D^0 +D^- +A \dot X -A^- \dot X ,
\end{array}
\eeq
where we have introduced the off-diagonal matrix $A$ and diagonal matrices
$D^0$, $D^{\pm}$ by their matrix elements
$$
\begin{array}{l}
A_{jk}=(1-\delta_{jk})\Phi (x_j-x_k), 
\\ \\
\displaystyle{D^0_{jk}=\delta_{jk}\sum_{l\neq j}\dot x_l\zeta (x_j -x_l)},
\quad
\displaystyle{D^{\pm}_{jk}=\delta_{jk}\sum_{l}\dot x_l\zeta (x_j -x_l\pm \eta )}.
\end{array}
$$
Taking into account that ${\bf c}^{*T}A^-=k{\bf c}^{*T}\dot X^{-1}$,
$A^-{\bf c}=k\dot X^{-1}{\bf c}$, we have:
$$
{\bf c}^{*T}
\Bigl (M^* E_i A^- -E_i A^- M -E_i \dot A^-\Bigr ){\bf c}
$$
$$
={\bf c}^{*T}\Bigl (k(D^--D^0)E_i\dot X^{-1}+kAE_i +E_iA^-(D^+-D^0)
$$
$$\phantom{aaaaaaaaaaaaaa}-E_i
(A^-\dot XA -A\dot X A^-)-kE_iA -E_i \dot XB^-+E_iB^- \dot X\Bigr ){\bf c}
$$
Using the identities (\ref{id1}), (\ref{id2}), one can see after some
calculations that the result is
\beq\label{pt15}
{\bf c}^{*T}
\Bigl (M^* E_i A^- -E_i A^- M -E_i \dot A^-\Bigr ){\bf c}=
k{\bf c}^{*T}[A, E_i]{\bf c}.
\eeq
It is easy to see that
$$
\dot x_i^{-1}{\bf c}^{*T}k[A, E_i]{\bf c}={\bf c}^{*T}\Bigl (
AE_iA^{-}-A^- E_i A\Bigr ) {\bf c}.
$$
The identities (\ref{id1}), (\ref{id2}) allow us to prove the relation
$$
AE_iA^{-}-A^- E_i A 
$$
$$
=[E_i, B^-]
+\sum_{l\neq i} \zeta (x_l-x_i)E_l A^- -
\sum_{l} \zeta (x_l-x_i+\eta ) A^-E_l
$$
$$
\phantom{aaaaaaaaaaaaaaaaaaaaaa}+\sum_{l\neq i}\zeta (x_l-x_i)A^-E_l
-\sum_{l} \zeta (x_l-x_i-\eta ) E_l A^-.
$$
Therefore, we have
\beq\label{pt16}
\begin{array}{l}
\displaystyle{
\dot x_i^{-1}\p_{t_m}\dot x_i =\res_{\infty}\left (z^{m}k^{-1}{\bf c}^{*T}
\Bigl ([E_i, B^-]+2 \sum_{l\neq i}\zeta (x_l-x_i)E_l A^-\right.}
\\ \\
\displaystyle{\phantom{aaaaaaaaaaaaa}\left. -\sum_{l} \zeta (x_l-x_i+\eta ) A^-E_l-
\sum_{l} \zeta (x_l-x_i-\eta ) A^- E_l \Bigr ){\bf c}\right ).}
\end{array}
\eeq

Now let us transform the rest part of (\ref{pt12a}):
$$
-\sum_{k\neq i}
(\p_{t_m}x_i-\p_{t_m}x_k)(\zeta (x_i-x_k+\eta )-\zeta (x_i-x_k))
$$
$$
=\res_{\infty}\left (z^{m}k^{-1}{\bf c}^{*T}\Bigl (
A^- (\tilde D^+\! -\! \tilde D^0)E_i -\! \sum_{l} \zeta (x_i-x_l+\eta ) A^-E_l
+\!\sum_{l\neq i}\zeta (x_i-x_l) A^-E_l\Bigr ){\bf c}\right )
$$
$$
=\res_{\infty}\left (z^{m}{\bf c}^{*T}\Bigl (\dot X^{-1}
(\tilde D^+ - \tilde D^0)E_i \vphantom{\sum_l^k}\right.
$$\hspace{-4cm}
$$
\left.
\phantom{aaaaaaaaaaaaaaaaaa}-\! \dot X^{-1}\sum_{l} \zeta (x_i-x_l+\eta ) E_l
+\!\dot X^{-1} \sum_{l\neq i}\zeta (x_i-x_l) E_l\Bigr ){\bf c}\right ).
$$
Collecting everything together, we finally find:
$$\hspace{-4cm}
\p_{t_m}p_i=\res_{\infty}\left (z^{m}k^{-1}{\bf c}^{*T}\Bigl (
k\dot X^{-1}
(\tilde D^+ - \tilde D^0)E_i \vphantom{\sum_l^k}\right.
$$
$$
\phantom{aaaaaaaaaaaaaaaaaa}- k\dot X^{-1}\sum_{l} \zeta (x_i-x_l+\eta ) E_l
+k\dot X^{-1} \sum_{l\neq i}\zeta (x_i-x_l) E_l
$$
$$\hspace{-3cm}
+[E_i, B^-]+2k\dot X^{-1}\sum_{l\neq i}E_l \zeta (x_l-x_i)
$$
$$\left. \phantom{aaaaaaaaaaaaaaaaaa}
-
k\dot X^{-1}\sum_{l}E_l \zeta (x_l-x_i+\eta )-k\dot X^{-1}\sum_{l}E_l \zeta (x_l-x_i-\eta )
\Bigr ){\bf c}\right )
$$
$$\hspace{-4cm}
=\res_{\infty}\left (z^{m}k^{-1}{\bf c}^{*T}\Bigl (
[E_i, B^-]+k\dot X^{-1}
(\tilde D^+ - \tilde D^0)E_i \vphantom{\sum_l^k}\right.
$$
$$\left. \phantom{aaaaaaaaaaaaaaaa}
+k\dot X^{-1} \sum_{l\neq i}\zeta (x_l-x_i) E_l-k\dot X^{-1} \sum_{l}\zeta (x_l-x_i+\eta )
E_l \Bigr ){\bf c}\right )
$$
$$\hspace{-4cm}
=\res_{\infty}\left (z^{m}k^{-1}{\bf c}^{*T}\Bigl (
[E_i, B^-]+\dot X^{-1}A^-
(\tilde D^+ - \tilde D^0)E_i \vphantom{\sum_l^k}\right.
$$
$$\left. \phantom{aaaaaaaaaaaaaaaa}
+\sum_{l\neq i}\zeta (x_l-x_i) E_lA^--\sum_{l}\zeta (x_l-x_i+\eta )
E_lA^- \Bigr ){\bf c}\right )
$$
$$\hspace{-9cm}
=\res_{\infty}\Bigl (z^{m}k^{-1}{\bf c}^{*T}Y_i {\bf c}\Bigr ),
$$
where $Y_i$ is the right hand side of (\ref{pt9}). Therefore, the second half of the 
Hamiltonian equations is proved. 

Let us comment on how the result will be modified if we consider a more general tau-function
then (\ref{tauell}) of the form
\beq\label{pt17}
\tau (x, {\bf t})=e^{Q(x, {\bf t})}\prod_{i=1}^N \sigma (x-x_i({\bf t})),
\eeq
where
\beq\label{pt18}
Q(x, {\bf t})=cx^2 +x\sum_{j\geq 1}a_j t_j +b({\bf t})
\eeq
with some constants $c$, $a_j$. (Here we have put $\bar {\bf t}=0$ for simplicity.)
Repeating the arguments leading to (\ref{e7}), one can see that now the
first equation in  (\ref{e7}) will be modified as
\beq\label{pt19}
k=ze^{\eta \zeta (\lambda )-\alpha (z)}, \qquad
\alpha (z)=\sum_{j\geq 1}\frac{a_j}{j}\, z^{-j}.
\eeq
Instead of (\ref{pt6}) we will have
\beq\label{pt20}
\p_{p_i}\log k = -\, \frac{\p_{p_i}\lambda (z)}{z\lambda '(z)}\, (1-z\alpha '(z)),
\eeq
so the Hamiltonian for the $m$-th flow will be a linear combination of $H_m$ and $H_j$ with
$1\leq j<m$. 

\subsection{Negative times}

First of all, let us obtain the relation between the velocities $\dot x_i=\p_{t_1}x_i$ and
$x_i'=\p_{\bar t_1}x_i$. The relation follows from the Toda equation (\ref{mkp18a}), where
we substitute the tau-function (\ref{tauell}) for elliptic solutions:
$$
\sum_i \dot x_i' \zeta (x-x_i)-\sum_i \dot x_i x_i' \wp (x-x_i)=1-
\prod_i \frac{\sigma (x-x_i+\eta )\sigma (x-x_i-\eta )}{\sigma ^2(x-x_i)}.
$$
Equating the coefficients in front of the second order poles, we get
\beq\label{nt1}
\dot x_i x_i'=-\sigma ^2(\eta )\prod_{j\neq i}
\frac{\sigma (x_i-x_j+\eta )\sigma (x_i-x_j-\eta )}{\sigma ^2(x_i-x_j)}.
\eeq

Our strategy is to solve the linear problems (\ref{mkp20c}) for the complimentary
wave functions $\phi$, $\phi^{\dag}$ represented as linear combinations of the elementary
double-Bloch functions as
\beq\label{nt2}
\begin{array}{l}
\displaystyle{
\phi (x)=\tilde k^{x/\eta}e^{\xi ({\bf t}, z^{-1})}\sum_i b_i \Phi (x-x_i+\eta , -\mu )},
\\ \\
\displaystyle{
\phi^{\dag} (x)=\tilde 
k^{-x/\eta}e^{-\xi ({\bf t}, z^{-1})}\sum_i b_i^* \Phi (x-x_i-\eta , \mu )},
\end{array}
\eeq
where $\mu , \tilde k$ are new spectral parameters to be 
connected with $\lambda , k$ later and $b_i$, $b_i^*$ do not depend on $x$. 
Identifying the monodromy properties of the functions (\ref{nt2}) and 
(\ref{mkp19b}), (\ref{mkp20b}), we obtain 
the relations between $\tilde k, z$ and $\mu$ in the same way 
as relations (\ref{e7}), (\ref{e8}) were obtained:
\beq\label{nt3}
\left \{\begin{array}{l}
\tilde k=z\, e^{-\eta\zeta (\mu )},
\\ \\
\displaystyle{\mu = -\sum_{m\geq 1}\frac{z^m}{m}\, \p_{\bar t_m}\! \sum_ix_i.}
\end{array} \right.
\eeq
It then follows that
\beq\label{nt3a}
\tilde k =k e^{-\eta \zeta (\lambda )-\eta \zeta (\mu )}.
\eeq
We note that the spectral parameters $\tilde k$ and $z$ here have
a different meaning than in (\ref{e7}), (\ref{e8}): they are local parameters on the spectral
curve in the vicinity of the point $P_0$ while in (\ref{e7}), (\ref{e8}) $k^{-1}$, $z^{-1}$
are local parameters 
in the vicinity of the point $P_\infty$. 

The substitution of (\ref{nt2}) into (\ref{mkp20c}) with
$$
\bar v(x)=\sum_i x_i' \Bigl (\zeta (x-x_i)-\zeta (x-x_i+\eta )\Bigr )
$$
gives, after cancellation of poles at $x=x_i$, the following conditions:
\beq\label{nt4}
\begin{array}{l}
\displaystyle{
x_i'\sum_j b_j \Phi (x_i-x_j +\eta , -\mu )=\tilde k^{-1}b_i,}
\\ \\
\displaystyle{
x_i'\sum_j b_j^* \Phi (x_i-x_j -\eta , \mu )=-\tilde k^{-1}b_i^*.}
\end{array}
\eeq
Using the identity (\ref{id0}), we can write them in the matrix form as follows:
\beq\label{nt5}
{\bf b}^TX'^{-1}\bar L(\mu )X'=-\tilde k^{-1}{\bf b}^T, \qquad
\bar L(\mu ){\bf b}^*=-\tilde k^{-1}{\bf b}^*.
\eeq
Here the matrix $\bar L(\mu)$ is
\beq\label{nt6}
\bar L_{ij}(\mu)=x_i'\Phi (x_i-x_j-\eta , \mu )=
-\sigma ^2(\eta )e^{-p_i}\Phi (x_i-x_j-\eta, \mu )\prod_{l\neq i}
\frac{\sigma (x_i-x_l -\eta )}{\sigma (x_i-x_l)},
\eeq
where we have used the relation (\ref{nt1}). 

Equations (\ref{nt5}) allow one to write the equation of the spectral curve
in the form
\beq\label{nt7}
\det \Bigl (\tilde k^{-1}I+\bar L(\mu )\Bigr )=0.
\eeq
Let us show that it is the same spectral curve as the one given by (\ref{spec1})
and discussed in section \ref{section:curve}. To show this, we find the inverse of the Lax
matrix $L(\lambda )$ given by (\ref{ell7}). We have  
$(L^{-1}(\lambda ))_{kl}=
(-1)^{k+l}\mbox{minor}\, {}_{lk}/\det L(\lambda )$ and use the fact that both numerator
and denominator here are determinants of the elliptic Cauchy matrices given explicitly
by (\ref{spec4}). In particular,
\beq\label{nt8}
\det L(\lambda )=e^{N\eta \zeta (\lambda )}\Bigl (\prod_i \dot x_i\Bigr )
\frac{\sigma (\lambda -N\eta )}{\sigma (\lambda )\sigma^N(-\eta )}\prod_{i\neq j}
\frac{\sigma (x_i-x_j)}{\sigma (x_i-x_j-\eta )}.
\eeq
After some simple transformations, we obtain for the $(L^{-1}(\lambda ))_{kl}$:
\beq\label{nt9}
\begin{array}{c}
\displaystyle{
(L^{-1}(\lambda ))_{kl}=-e^{(x_l-x_k-\eta )\zeta (\lambda )}x_l'\, 
\frac{\sigma (N\eta -\lambda +x_l-x_k-\eta )}{\sigma (N\eta -\lambda )\sigma (x_l-x_k-\eta )}}
\\ \\
\displaystyle{\times
\prod_{i\neq l}\frac{\sigma (x_l-x_i)}{\sigma (x_l-x_i+\eta )}
\prod_{j\neq k}\frac{\sigma (x_k-x_j+\eta )}{\sigma (x_k-x_j)}.}
\end{array}
\eeq
Therefore, we see that
\beq\label{nt10}
e^{\eta \zeta (\lambda )}(L^T(\lambda ))^{-1}\cong -e^{-\eta \zeta (\mu )}
\bar L(\mu ), \qquad \mu =N\eta -\lambda ,
\eeq
where $\cong$ means equality up to a similarity transformation. Taking into account
(\ref{nt10}), (\ref{nt3a}), 
we write the spectral curve (\ref{nt7}) in the form
$$
\det \Bigl (k^{-1}I-L^{-1}(\lambda )\Bigr )=0
$$
which is the same as (\ref{spec1}). However, in the previous sections and here we
expand the spectral curve in the vicinity of different points: it was the point
$P_{\infty}=(\infty , 0)$ in the previous sections and $P_0=(0, N\eta )$ here. 

Using equation (\ref{mkp22a}) and repeating the calculations leading to (\ref{pt3}),
(\ref{pt5}), we obtain the relations
\beq\label{nt11}
\p_{\bar t_m}x_i=-\res_0 \Bigl (z^{-m}\tilde k^{-1}{\bf b}^T X'^{-1} \p_{p_i}\bar L(\mu )
{\bf b}^*\Bigr )
\eeq
and
\beq\label{nt12}
{\bf b}^T X'^{-1}{\bf b}^* =\tilde k^2\mu '(z).
\eeq
A similar chain of equalities as the one after (\ref{pt5}) leads to 
\beq\label{nt13}
\p_{\bar t_m}x_i=\res_0 \Bigl (z^{-m-1}\p_{p_i}\mu (z)\Bigr )=\frac{\p \bar H_m}{\p p_i},
\eeq
where
\beq\label{nt14}
\bar H_m =\res_0 \Bigl (z^{-m-1}\mu (z)\Bigr )=-\res_0 \Bigl (z^{-m-1}\lambda (z)\Bigr ).
\eeq
We see that the Hamiltonians for the negative time flows are obtained from the expansion
of $\lambda (z)$ as $z\to 0$:
\beq\label{nt14a}
\lambda (z)=N\eta -\sum_{m\geq 1}\bar H_m z^m.
\eeq

The other half of the Hamiltonian equations,
\beq\label{nt15}
\p_{\bar t_m}p_i=-\frac{\p \bar H_m}{\p x_i},
\eeq
can be proved in the same way as equations (\ref{pt10}) were proved. 

\section{Degenerations of elliptic solutions}

\subsection{Rational limit}

In the rational limit $\omega , \omega ' \to \infty$ $\sigma (x)=x$ and
\beq\label{rat1}
L_{ij}(\lambda )=\left (\frac{\dot x_i}{x_i\! -\! x_j\! -\! \eta}+\frac{\dot x_i}{\lambda}
\right )e^{-(x_i-x_j)/\lambda +\eta /\lambda},
\eeq
where
$$
\dot x_i=e^{p_i}\prod_{l\neq i}\frac{x_i-x_l+\eta}{x_i-x_l}.
$$
The equation of the spectral curve is
$$
\det \Bigl (kI-e^{\eta /\lambda}(L_{\rm rat}+\lambda^{-1}\dot X E)\Bigr )=0,
$$
where $L_{\rm rat}$ is the Lax matrix of the rational Ruijsenaars-Schneider system
with matrix elements
\beq\label{rat2}
(L_{\rm rat})_{ij}=\frac{\dot x_i}{x_i\! -\! x_j\! -\! \eta}.
\eeq
Recalling the connection between the spectral parameters $k, z, \lambda$ 
$k=ze^{\eta /\lambda}$, we can write the equation of the spectral curve in the form
\beq\label{rat3}
\det \Bigl (zI-L_{\rm rat}-\lambda^{-1}\dot X E\Bigr )=0.
\eeq
Since $E$ is the rank 1 matrix, we have
$$
\det \left (I-\lambda^{-1}\dot XE\, \frac{1}{zI-L_{\rm rat}}\right )=
1-\lambda^{-1}\mbox{tr}\, \Bigl (\dot XE\, \frac{1}{zI-L_{\rm rat}}\Bigr )=0,
$$
so
\beq\label{rat4}
\lambda =\mbox{tr}\, \Bigl (\dot XE\, \frac{1}{zI-L_{\rm rat}}\Bigr ).
\eeq
As $z\to \infty$ we expand this as
\beq\label{rat4a}
\lambda (z)=\sum_{m\geq 1}z^{-m}\mbox{tr}\, \Bigl (\dot X EL_{\rm rat}^{m-1}\Bigr ).
\eeq
It is easy to check the commutation relation
\beq\label{rat5}
XL_{\rm rat}-L_{\rm rat}X=\dot XE +\eta L_{\rm rat}.
\eeq
Using it, we have
$$
\mbox{tr}\, \Bigl (\dot X EL_{\rm rat}^{m-1}\Bigr )=\mbox{tr}\Bigl (
XL_{\rm rat}^m - L_{\rm rat}XL_{\rm rat}^{m-1}-\eta L_{\rm rat}^m\Bigr )=-\eta \,
\mbox{tr}\, L_{\rm rat}^m.
$$
Therefore, we conclude that
\beq\label{rat6}
\lambda (z)=-\eta \sum_{m\geq 1}z^{-m}\mbox{tr}\, L_{\rm rat}^m \quad \mbox{as $z\to \infty$}
\eeq
and thus the Hamiltonians for positive time flows are
\beq\label{rat7}
H_m=-\eta \mbox{tr}\, L_{\rm rat}^m.
\eeq
This agrees with the result of papers \cite{Iliev07,Z14}. 

In order to find the Hamiltonians for negative time flows we expand (\ref{rat4})
as $z\to 0$:
\beq\label{rat8}
\lambda (z)=-\sum_{m\geq 0}z^m \mbox{tr}\, (\dot XEL_{\rm rat}^{-m-1})=
N\eta +\eta \sum_{m\geq 1}z^m \mbox{tr}\, L_{\rm rat}^{-m}.
\eeq
Therefore,
\beq\label{rat7a}
\bar H_m=-\eta \, \mbox{tr}\, L_{\rm rat}^{-m}.
\eeq
This agrees with the result of the paper \cite{PZ19}. 

\subsection{Trigonometric limit}

We now pass to the trigonometric limit. Let $\pi i/\gamma$ be period of the 
trigonometric (or hyperbolic) functions with the second period tending to infinity.
The Weierstrass functions in this limit are
$$
\sigma (x)=\gamma^{-1}e^{-\frac{1}{6}\, \gamma^2x^2}\sinh (\gamma x),
\quad
\zeta (x)=\gamma \coth (\gamma x)-\frac{1}{3}\, \gamma^2 x.
$$
The tau-function for trigonometric solutions is
\beq\label{trig1}
\tau =\prod_{i=1}^N \Bigl (e^{2\gamma x}-e^{2\gamma x_i}\Bigr ),
\eeq
so we should consider
\beq\label{trig2}
\tau =\prod_{i=1}^N \sigma (x-x_i)e^{\frac{1}{6}\, \gamma^2
(x-x_i)^2 +\gamma (x+x_i)}.
\eeq
Similarly to the KP case \cite{PZ21},
equation (\ref{e7}) with this choice acquires the form
\beq\label{trig3}
\log k=\log z+\eta \gamma \coth (\gamma \lambda ).
\eeq

The trigonometric limit of the function $\Phi (x, \lambda )$ is
$$
\Phi (x, \lambda )=\gamma \Bigl (\coth (\gamma x)+\coth (\gamma \lambda )\Bigr )
e^{-\gamma x \coth (\gamma \lambda )}
$$
and $L(\lambda )$ takes the form
$$
L_{ij}(\lambda )=\gamma e^{\eta \gamma \coth (\gamma \lambda )}
e^{-\gamma \coth (\gamma \lambda )(x_i-x_j)}\Bigl (\dot x_i\coth (\gamma (
x_i\! -\! x_j\! -\! \eta ) )+\dot x_i \coth (\gamma \lambda )\Bigr ).
$$
For further calculations it is convenient to change the variables as
\beq\label{trig4}
w_i=e^{2\gamma x_i}, \qquad q=e^{2\gamma \eta}
\eeq
and introduce the diagonal matrix $W=\mbox{diag}\, (w_1, w_2, \ldots , w_N)$. 
In this notation, the equation of the spectral curve acquires the form
\beq\label{trig5}
\det \Bigl (zI-q^{-1/2}W^{1/2}L_{\rm trig}W^{-1/2}-\gamma (\coth (\gamma \lambda )-1)
\dot X E\Bigr )=0,
\eeq
where $L_{\rm trig}$ is the Lax matrix of the trigonometric Ruijsenaars-Schneider model:
\beq\label{trig6}
(L_{\rm trig})_{ij}=2\gamma q^{1/2}\, \frac{\dot x_i w_i^{1/2}w_j^{1/2}}{w_i-qw_j}
\eeq
(see \cite{PZ19}). Again, using the fact that $E$ is the rank 1 matrix, we obtain
from (\ref{trig5})
$$
(\coth (\gamma \lambda )-1)^{-1}=\gamma \, \mbox{tr} \left (
\dot XE\, \frac{1}{zI-q^{-1/2}W^{1/2}L_{\rm trig}W^{-1/2}}\right )
$$
or
$$
\lambda =\frac{1}{2\gamma}\, \log \left (1+2\gamma \, \mbox{tr}\Bigl (
\dot XE\, \frac{1}{zI-q^{-1/2}W^{1/2}L_{\rm trig}W^{-1/2}}\Bigr )\right ).
$$
Applying the formula $\det (I+Z)=1+\mbox{tr}\, Z$ for any matrix $Z$ of rank 1 in the
opposite direction, we have
\beq\label{trig7}
\lambda =\frac{1}{2\gamma}\, \log \det \left (
\Bigl (zI-q^{-1/2}W^{1/2}L_{\rm trig}W^{-1/2}+2\gamma \dot XE\Bigr )
\frac{1}{zI-q^{-1/2}W^{1/2}L_{\rm trig}W^{-1/2}}\right ).
\eeq
Next, we use the trigonometric analogue of the relation (\ref{rat5}):
\beq\label{trig8}
2\gamma \dot XE=q^{-1/2}W^{1/2}L_{\rm trig}W^{-1/2}-q^{1/2}
W^{-1/2}L_{\rm trig}W^{1/2}
\eeq
which can be easily checked. Using this relation, we obtain from (\ref{trig7}):
$$
\lambda =\frac{1}{2\gamma}\, \log \det \Bigl (zI-q^{1/2}W^{-1/2}L_{\rm trig}W^{1/2}\Bigr )-
\frac{1}{2\gamma}\, \log \det \Bigl (zI-q^{-1/2}W^{1/2}L_{\rm trig}W^{-1/2}\Bigr )
$$
$$
=\frac{1}{2\gamma}\, \log \det 
\frac{1-z^{-1}q^{1/2}L_{\rm trig}}{1-z^{-1}q^{-1/2}L_{\rm trig}}=-
\sum_{m\geq 1}z^{-m}\frac{q^{m/2}-q^{-m/2}}{2\gamma m}\, \mbox{tr}\, L_{\rm trig}^m.
$$
Therefore, we finally obtain
\beq\label{trig9}
\lambda (z)=-\sum_{m\geq 1}z^{-m}\, \frac{\sinh (\gamma \eta m)}{\gamma m}\, 
\mbox{tr}\, L_{\rm trig}^m
\eeq
and so the Hamiltonians are
\beq\label{trig10}
H_m= -\frac{\sinh (\gamma \eta m)}{\gamma m}\, 
\mbox{tr}\, L_{\rm trig}^m.
\eeq
This agrees with the result of paper \cite{PZ19}.

\section{Examples of the Hamiltonians}

We return to the general elliptic case. Let us introduce renormalized integrals of motion
\beq\label{eh1}
J_m=\frac{\sigma (m\eta )}{\sigma ^m(\eta )}\, I_m, \qquad m=\pm 1, \pm 2, \ldots , \pm N,
\eeq
where $I_m$ are integrals of motion (\ref{intr1}), (\ref{intr1a}). The equation 
of the spectral curve (\ref{spec5}) is
\beq\label{eh2}
z^N +\sum_{n=1}^N \phi_n(\lambda )J_n z^{N-n}=0,
\eeq
where
\beq\label{eh3}
\phi_n(\lambda )=\frac{\sigma (\lambda -n\eta )}{\sigma (\lambda )\sigma (n\eta )}.
\eeq
It can be expanded as $\lambda \to 0$ ($z\to \infty$) as follows:
$$
\phi_n(\lambda )=-\frac{1}{\sigma (\lambda )}+\zeta (n\eta )\, 
\frac{\lambda}{\sigma (\lambda )}-(\zeta^2(n\eta )-\wp (n\eta ))\,
\frac{\lambda^2}{2\sigma (\lambda )}+O(z^{-2}).
$$
Expanding the equation of the spectral curve, we have:
$$
1+\zeta (\eta )J_1 z^{-1} -\frac{1}{2}\, (\zeta^2(\eta )-\wp (\eta ))J_1^2 z^{-2}
+\zeta (2\eta )J_2 z^{-2}=\frac{1}{\lambda z}\Bigl (J_1+J_2 z^{-1}+J_3 z^{-2}\Bigr )
+O(z^{-3})
$$
or
$$
\lambda (z)=\frac{z^{-1}(J_1+J_2 z^{-1}+J_3 z^{-2}
+O(z^{-3}))}{1+\zeta (\eta )J_1 z^{-1} 
+\zeta (2\eta )J_2 z^{-2} -\frac{1}{2}\, (\zeta^2(\eta )-\wp (\eta ))J_1^2 z^{-2}
+O(z^{-3})}.
$$
Expanding this in the series $\displaystyle{\lambda (z)=\sum_{m\geq 1}H_m z^{-m}}$, we 
find the first three Hamiltonians:
\beq\label{eh4}
\begin{array}{l}
H_1=J_1,
\\ \\
H_2=J_2-\zeta (\eta )J_1^2,
\\ \\
H_3=J_3-(\zeta (\eta )+\zeta (2\eta ))J_1J_2 +\Bigl (\frac{3}{2}\zeta^2(\eta )-\frac{1}{2}
\wp (\eta )\Bigr )J_1^3.
\end{array}
\eeq

In the rational limit one can obtain a general formula for $H_m$. In this limit
$$
\phi_n(\lambda)=\frac{1}{n\eta }-\frac{1}{\lambda}
$$
and the equation of the spectral curve becomes
\beq\label{eh5}
z^N +\frac{1}{\eta}\sum_{n=1}^N \frac{J_n}{n}z^{N-n}=\frac{1}{\lambda}
\sum_{n=1}^N J_n z^{N-n},
\eeq
or, recalling that in the rational limit $J_n=n\eta^{1-n}I_n$,
\beq\label{eh6}
\lambda(z)= \frac{\eta \sum\limits_{n=1}^{N}nI_n(\eta z)^{-n}}{1+
\sum\limits_{n=1}^{N}I_n(\eta z)^{-n}}.
\eeq
Expanding the right hand side in powers of $z^{-1}$, we obtain:
\beq\label{eh7}
\begin{array}{l}
H_1=I_1,
\\ \\
H_2=\eta^{-1}(2I_2-I_1^2),
\\ \\
H_3=\eta^{-2}(3I_3-3I_1I_2 +I_1^3).
\end{array}
\eeq
In general, from (\ref{eh6}) it follows that
\beq\label{eh8}
H_m=(-\eta )^{1-m}\sum_{|\nu |=m}C_{\nu}^{m}I_{\nu_1}\ldots I_{\nu_{\ell}},
\eeq
where the sum is taken over all Young diagrams $\nu$ having $|\nu |=m$ boxes with
non-zero rows $\nu_1, \ldots , \nu_{\ell}$ and $C_{\nu}^m$ is the matrix of transition from
the basis of elementary symmetric polynomials $e_j$ to power sums $p_k$. Indeed, it is easy to see
that the equality
$$
\sum_{n=1}^N (-1)^{n-1}p_n z^{-n}=
\frac{\sum\limits_{m=1}^N me_mz^{N-m}}{\sum\limits_{r=0}^N e_rz^{N-r}}
$$
is equivalent to the well known Newton's identity 
$$
me_m=\sum_{n=1}^m (-1)^{n-1}p_ne_{m-n}
$$
for the symmetric functions. The explicit formula is
\beq\label{eh9}
H_m=-m\eta^{1-m}\!\!\!\! \sum_{{\scriptsize \begin{array}{c}r_1+2r_2+\ldots +mr_m=m
\\ r_1\geq 0, \ldots , r_m\geq 0
\end{array}}}\!\!\!\!
\frac{(r_1+\ldots +r_m -1)!}{r_1!\ldots r_m!}\, 
\prod_{i=1}^m (-I_i)^{r_i}.
\eeq

Let us now consider Hamiltonians for the negative time flows. The equation of the spectral curve
(\ref{spec5}) can be rewritten in the form
\beq\label{eh10}
\varphi_N (\lambda )+\sum_{n=1}^N \varphi_{N-n}(\lambda )I_{-n}z^n=0
\eeq
or, equivalently,
\beq\label{eh11}
\sigma (\mu )+\sum_{n=1}^N \frac{\sigma (n\eta -\mu )}{\sigma (n\eta )}\, J_{-n}z^n=0,
\quad \mu\equiv N\eta -\lambda .
\eeq
Expanding $\mu \to 0$ in powers of $z$ as $\displaystyle{\mu (z)=\sum_{m\geq 1}\bar H_mz^m}$,
we obtain:
\beq\label{eh12}
\begin{array}{l}
\bar H_1=-J_{-1},
\\ \\
\bar H_2=-J_{-2}-\zeta (\eta )J_{-1}^2,
\\ \\
\bar H_3=-J_{-3}-(\zeta (\eta )+\zeta (2\eta ))J_{-1}J_{-2} -
\Bigl (\frac{3}{2}\zeta^2(\eta )-\frac{1}{2}
\wp (\eta )\Bigr )J_{-1}^3.
\end{array}
\eeq
In the rational limit we have
\beq\label{eh13}
\begin{array}{l}
\bar H_1=\eta^2 I_{-1},
\\ \\
\bar H_2=\eta^3(2I_{-2}-I_{-1}^2),
\\ \\
\bar H_3=\eta^4(3I_{-3}-3I_{-1}I_{-2} +
I_{-1}^3).
\end{array}
\eeq
One can see that the properly arranged limit $\eta \to 0$ in (\ref{eh7}) of the Hamiltonians 
for positive time flows yields
Hamiltonians of the Calogero-Moser model while the Hamiltonians (\ref{eh13}) for negative
time flows disappear in this limit.

\section{Conclusion}

The main result of this paper is establishing the precise correspondence between
elliptic solutions of the 2D Toda lattice hierarchy and the hierarchy of the Hamiltonian
equations for the integrable elliptic Ruijsenaars-Schneider model with higher Hamiltonians. 
The zeros of the tau-function move as particles of the Ruijsenaars-Schneider model.
We have shown that the $m$th time flow $t_m$ of the 2DTL hierarchy gives rise to the 
flow with the Hamiltonian $H_m$ of the Ruijsenars-Schneider model which is obtained
as $m$th coefficient of the expansion of the spectral curve $\lambda (z)$ 
in negative powers of $z$ as $z\to \infty$. Moreover, the 
$m$th time flow $\bar t_m$ of the 2DTL hierarchy corresponds to 
the flow with the Hamiltonian $\bar H_m$ of the Ruijsenars-Schneider model which is obtained
as $m$th coefficient of the expansion of the spectral curve $\lambda (z)$ 
in positive powers of $z$ as $z\to 0$. The first few Hamiltonians 
were found explicitly. 

For rational and trigonometric degenerations
of elliptic solutions the previous results of the papers \cite{Iliev07,Z14,PZ19}
obtained there by a different method are reproduced:
the $m$th time flow $t_m$ of the 2DTL hierarchy gives rise to the 
flow with the Hamiltonian $H_m$
proportional to $\mbox{tr}\, L^m$, where $L$ is the Lax matrix, while the time flow
$\bar t_m$ gives rise to the Hamiltonian flow with the Hamiltonian $\bar H_m$ 
proportional to $\mbox{tr}\, L^{-m}$. 

\section*{Acknowledgments}

The research of A.Z. has been funded within the framework of the
HSE University Basic Research Program and the Russian Academic Excellence Project '5-100'.

\addcontentsline{toc}{section}{\hspace{6mm}Acknowledgments}

\end{document}